\documentclass[%
 aip,
 amsmath,amssymb,
 reprint,%
]{revtex4-1}
\usepackage[main=english]{babel}
\usepackage{graphicx}
\usepackage{dcolumn}
\usepackage{bm}

\usepackage[utf8]{inputenc}
\usepackage[T1]{fontenc}
\usepackage{mathptmx}
\usepackage{etoolbox}

\usepackage{hyperref}
\usepackage{url} 
\usepackage{color}
\usepackage[ margin=1in]{geometry}
\usepackage{natbib}
\usepackage{amsmath}
\usepackage{braket}
\usepackage{bbm}

\makeatletter
\def\@email#1#2{%
 \endgroup
 \patchcmd{\titleblock@produce}
  {\frontmatter@RRAPformat}
  {\frontmatter@RRAPformat{\produce@RRAP{*#1\href{mailto:#2}{#2}}}\frontmatter@RRAPformat}
  {}{}
}%
\makeatother
\begin{document}

\preprint{AIP/123-QED}

\title[]{Intrinsic atomic calibration of oscillating magnetic fields in ULF and VLF bands}
\author{Zak Johnston}%
\author{Paul F. Griffin}%
\author{Erling Riis}
\author{Dominic Hunter}
\author{Marcin Mrozowski}
\author{Stuart J. Ingleby}%
 \email{zak.johnston@strath.ac.uk}
\affiliation{Department of Physics, SUPA, University of Strathclyde, Glasgow, G4 0NG, UK
}%

\date{\today}

\begin{abstract}
 We present a method for absolute calibration of received radio-frequency in the ultra low frequency (ULF), and very low frequency (VLF) range. This is achieved with the use of a radio frequency optically pumped magnetometer (RF-OPM). \textcolor{black}{We describe a method using an optically pumped sample where the RF broadening of the Cs magnetic resonance allows the magnitude of the received field to be calibrated against the ground-state gyromagnetic ratio of the Cs atoms.}
This frequency-based calibration avoids the geometric and electrostatic response functions that affect inductive sensors, such as fluxgates, search coils, and SQUID magnetometers. We demonstrate calibration of magnetic measurement using oscillating magnetic fields in the 300~Hz – 20~kHz range and a sensor noise floor of 15 fT.Hz$^{-1/2}$. This radio-frequency sensor may be used as a widely tunable narrowband receiver for communication, ranging, or penetrative conductivity imaging.
\end{abstract}

\maketitle

%

\section{\label{sec:level1}Introduction\protect}
The accurate and precise measurement of oscillating magnetic fields is paramount in many different sensing applications \cite{Hunter:22}. Radio frequency optically pumped magnetometers (RF-OPMs) have been utilised in applications that require non destructive testing; such as defect detection of materials \cite{s22249741,10.1063/5.0102402,10.1063/1.5083039}, or non invasive medical diagnosis \cite{Deans:2020isr}. These techniques use an induced eddy-current response to determine the material properties of the measured sample, this also allows for magnetic imaging of the sample from the received field, known as magnetic induction tomography \cite{Deans_2021}. RF-OPMs also have a high sensitivity and large tuning range; as such they can also be used for remote sensing applications such as communication and ranging in low signal amplitude and noisy environments \cite{MM}, such as communications through attenuating media. 
The detection of oscillating magnetic fields can be achieved using other types of magnetometers such as fluxgates or pick-up coils \cite{RIPKA20038}, all of which require calibration measurement of their generated and received oscillating magnetic fields. However, creating a defined calibration using standard coil geometries requires the manufacture of a coil with minimal defects in axial and transverse winding  \cite{523521}. We present an improved calibration method that is independent of these physical parameters.

RF-OPMs utilise a polarised ensemble of alkali-metal atoms in the vapour phase to measure magnetic fields \cite{PhysRevLett.95.063004, PhysRevApplied.11.044034}. An external magnetic field applied to an atomic system sets the ensemble's spin precession frequency, which provides information on the scalar or vector properties of the field \cite{S.Ingleby}. OPMs have inherent advantages when compared to other magnetometer devices, such as scalability \cite{Hunter:23}, sensitivity \cite{PhysRevLett.110.160802}, and non-cryogenic operation \cite{MU}.
RF-OPMs, utilising tuned resonant pickup of oscillating fields in lower-frequency bands, offer an improvement to inductive sensing, to which classical constraints between wavelength and receiver size are not applicable. Measurement of an oscillating field in the ULF and VLF frequency bands necessitates larger pick up coils as the sensitivity scales with coil size and sensitivity decreases with frequency \cite{Mooney_2017,SAVUKOV2007214}. 

Here we present a calibration method for the RF field in the ULF and VLF bands that exploits saturation of the atomic response to the received field. By application of a test RF field of varying magnitudes, we use the RF saturation contribution to the measured resonance lineshape to calibrate the amplitude of the applied field, which acts as a reference signal for other received magnetic signals. Eliminating the effect of bias-field variation on spin-exchange relaxation, we show that the measured field response follows the expected resonance lineshape, predicted by the optical Bloch equations. With this calibrated coil as a reference, we determine the sensor's intrinsic noise floor.

\section{Method and Apparatus}

Figure 1 depicts the experimental schematic of an RF-OPM, which operates on a double-resonance principle. This involves combining resonant laser light to drive optical transitions with ground-state magnetic resonances induced through optical pumping. 
\begin{figure}[hbt!]
\includegraphics[scale=0.43, angle = 0]{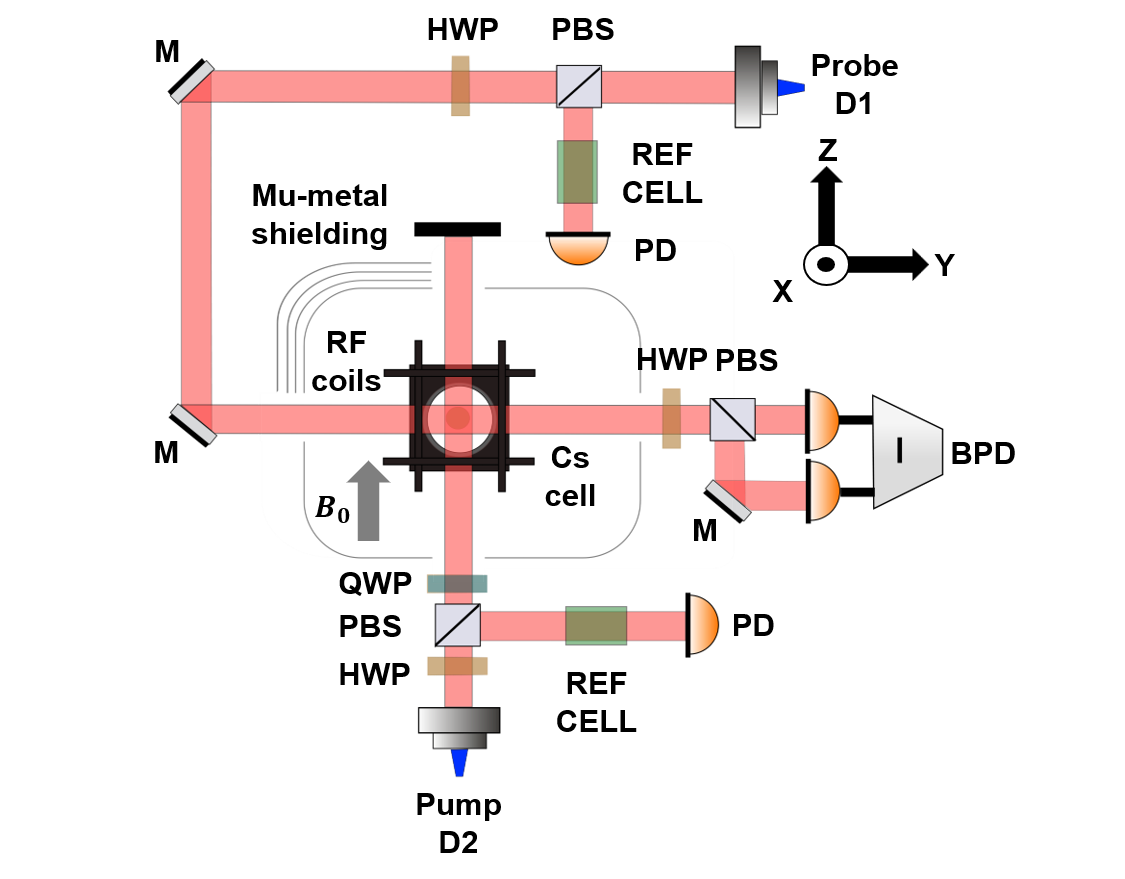}
\centering
\caption{Schematic of apparatus used. Double resonance magnetometer comprised of a volume holographic grating (VHG) pump and distributed Bragg reflector (DBR) probe laser, 4 layer mu-metal magnetic shielding which houses a 28~mm diameter glass blown paraffin coated cell containing a thermal vapour of caesium, internal RF modulation coil and internal static field generation coil. Two reference cells with their respective photodiode (PD), and a balanced polarimeter comprised of a half-wave plate (HWP), polarising beam splitter (PBS) and balanced photodetector (BPD).}
\end{figure}
A perpendicular two-beam configuration consisting of a pump and probe has been used to interrogate our caesium sample. Optical pumping is tuned to the D2 line ($852.34$~nm), where circularly polarised light drives induced $\sigma^{+}$ transitions of the  $F=3 \rightarrow F^{'} =4$ optical transition. The ground-state polarisation lifetime is maximised by the use of a paraffin anti-relaxation coating, allowing many wall collisions (up to $10^5$) before the spin destruction of the atomic polarisation has occurred \cite{Balabas:10, Castagna2009}. A static field $B_{0}$, parallel to the pump beam, has been generated using a low noise current driver \cite{mrozowski2022ultralow}, oriented along the z axis. The nonlinear Zeeman effect is negligible for low magnitudes of $B_{0}$, the degeneracy of the Zeeman states are lifted uniformly. Combined with efficient optical pumping, this generates an orientation moment in the system, where atoms begin to populate the $F=4$, $m_{f} = +4$ stretched state. The build-up of polarised atoms in the same state gives rise to a net magnetisation in the vapour cell, with the ensemble displaying a finite spin polarisation lifetime dictated by various spin relaxation mechanisms, mainly residual spin-destruction wall collisions. 

The ground states are probed on the D1 line ($894.59$~nm) using linearly polarised light red-detuned by 1.35~GHz from the $F=4 \rightarrow F^{'}= 3 $ transition. The spin orientation after the cell is measured with a balanced polarimeter, which determines the change in light polarisation before and after the cell through the measurement of Faraday rotation, where the degree of rotation is proportional to the transverse spin component \textcolor{black}{$M_{x}$.} 

\textcolor{black}{An RF magnetic field ($B_{RF}$) is applied perpendicular to the static field using a square, single-turn coil (5cm side length) orientated on the x-axis.} The change in spin orientation under the effect of both fields has been modeled semi-classically using the Bloch equations \cite{Bison:05}, and the steady-state solution in the rotating frame is found by considering the time derivative of the ground-state polarisation moment ${\bold {M}}$,

\begin{equation}
    \dot{M} = \begin{pmatrix}
\Gamma_{2} & \Delta & 0\\
-\Delta & \Gamma_{2} & \Omega_{RF} \\
0 & -\Omega_{RF} & \Gamma_{1}
\end{pmatrix}	
\begin{pmatrix}
M_{x} & \\
M_{y} & \\
M_{z}
\end{pmatrix}
+\Gamma_{1} \begin{pmatrix}
0 & \\
0 & \\
M_{0}
\end{pmatrix}
,
\end{equation}
\textcolor{black}{where $\Gamma_{1}$ is the longitudinal relaxation rate and $\Gamma_{2}$ the transverse relaxation rate,} $\Delta$ is the detuning from the Larmor frequency ($\Delta = \omega_{L} - \omega_{RF}$), $\Omega_{RF}$ is the magnetic Rabi frequency ($\Omega_{RF} = \frac{\gamma B_{RF}}{2} $), and $M_{0}$ is the steady state magnetisation.

The absorptive and dispersive line shapes, $M_{y}$ and $M_{x}$ respectively, have been measured by demodulation of the balanced polarimeter signal with reference to the phase and frequency of the applied field $B_{RF}$. By considering the steady-state solution to Eq. 1, the demodulated signals are fitted as a function of detuning from the Larmor frequency. \textcolor{black}{An example of a measured RF resonance with the associated model fit\cite{10.1063/1.4980159} is shown in Fig. 2.}

\begin{figure}[hbt!]
\includegraphics[scale=0.55]{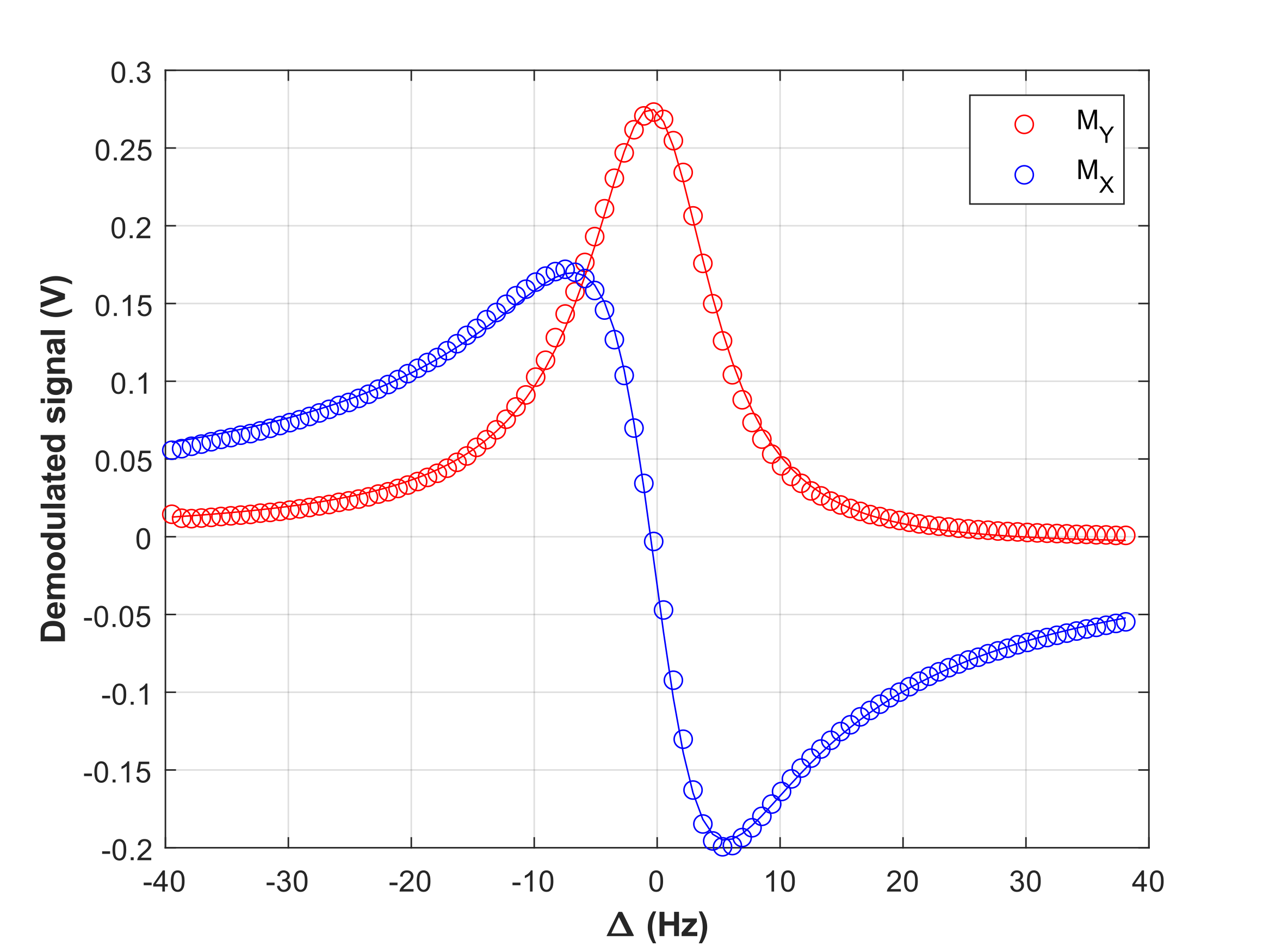}
\centering
\caption{Magnetic resonance generated from a frequency sweep of 100 samples over a detuning range of 80~Hz, for a total sampling time of 5 seconds. Both $M_{y}$ (red) and $M_{x}$ (blue) are fitted simultaneously with the best fit shown with solid lines, and the best fit parameters $M_{0}$ = 0.560~V, $\omega_{L}$ = 2.00~kHz, $\Gamma$ = 4.67~Hz, $\Omega_{RF}$ = 4.08~Hz.}
\end{figure}

\textcolor{black}{By examining how the magnetic resonance saturates as a function of the increased RF amplitude, we obtain our intrinsic calibration.} We can measure this atomic response to the field and determine the equivalent current through the relation to the species gyromagnetic ratio.

\section{Results}

\subsection{Calibration against RF saturation}

\textcolor{black}{It is convenient for the analysis to only consider the amplitude dependence of our signal and remove any dependence on phase, this removes the phase as a free parameter in the fitting model.} This was done by taking the quadrature sum of $M_{x}$ and $M_{y}$, $R = \sqrt{M^2_{X}+M^2_{Y}}$. The saturation of the RF resonance is induced by increasing the driving strength of the applied RF modulation. A voltage is applied across the RF coil in series with a fixed resistor. The data acquisition (DAQ) was used to synchronously digitize the polarimeter signal. 

An RF current range from 20~$\mu A$ to \textcolor{black}{160~$\mu A$} has been used, which has the equivalent field range 0.73~$nT$ to 5.80~$nT$, this allows for sufficient resonant response analysis across the non saturated and saturated regions. Saturation of the field-atom interaction occurs as the RF Rabi rate approaches the ground-state atomic relaxation rate, leading to a saturated state \cite{10.1063/5.0008273}. 
This saturation leads to a characteristic dip from the on-resonance response, decreasing the amplitude and increasing the width of the resonance, which can be seen in \textcolor{black}{Fig. 3.} The saturation parameter, $S$ is defined by the ratio of the magnetic Rabi rate $\Omega_{RF}$ and relaxation rate $\Gamma$. Saturation of the resonance occurs when $\frac{\Omega_{RF}}{\Gamma}$ > 1. The saturation dip begins after 60~$\mu A$ (2.19~$nT$). \textcolor{black}{At this value the saturation parameter was calculated to be 0.92, shown as the blue trace in Fig. 3.} 

\begin{figure}[hbt!]
\includegraphics[scale=0.55]{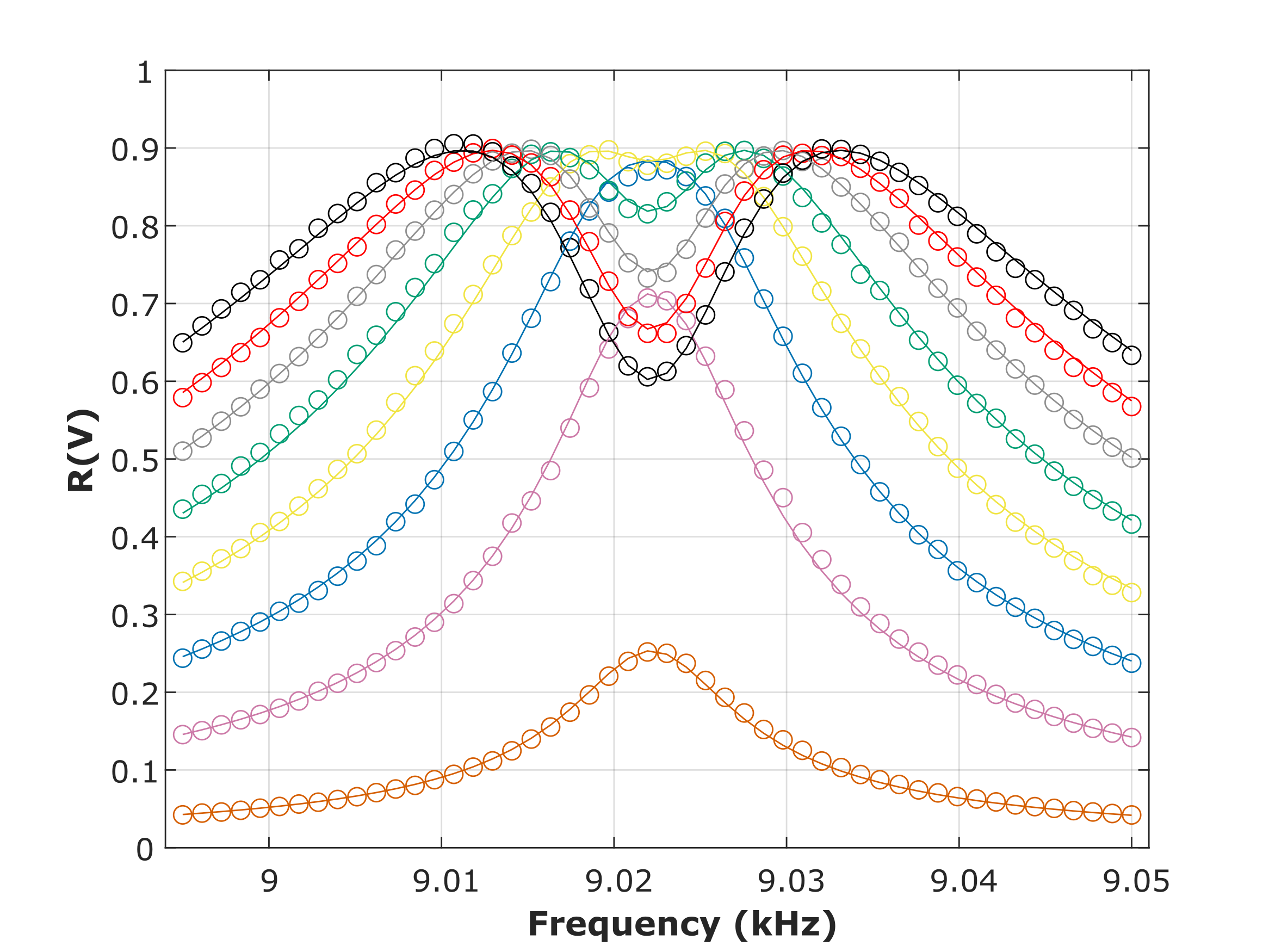}
\centering
\caption{Multiple R, resonant response to a swept RF frequency data with a global fit at a single Larmor frequency. Each resonant response represents the same detuning range across a central Larmor frequency with increasing RF drive from 0.73~$nT$ (orange) up to 5.80~$nT$ (black).} 
\end{figure}

\textcolor{black}{Analysis across the non-saturated and saturated region is necessary as the change in the resonance lineshape allows us to constrain the amplitude of the applied magnetic field.} 
We can relate the amplitude and width to the relaxation rate and Rabi rate to determine the magnitude of the applied RF field. The on-resonance response is given as

\begin{equation}
    M_{y}|_{\Delta = 0} = \frac{M_{0}\Omega_{RF}}{1+(\frac{\Omega_{RF}}{\Gamma})^2}.
\end{equation}

\textcolor{black}{ In the linear region the covariance between $M_{0}$ and $\Omega_{RF}$ is large and therefore cannot be constrained by curve fitting. In the saturated regime this covariance is decreased and the parameters can be fitted independently}, this emerges in the $\frac{\Omega_{RF}}{\Gamma}$ > 1 regime where we can infer information on the Rabi rate alone due to the broadening of the resonance. This then allows us to determine the affect that the saturated field has on the atomic response.

Figure 3 shows eight resonant responses to  a swept RF frequency, where each data set denotes a different \textcolor{black}{RF amplitude}. \textcolor{black}{As the RF field} is increased, the on-resonance RF amplitude follows until saturation where the line shape broadens. A global fit to all of these eight resonant responses is completed, in which the RF amplitude is parameterised by the applied RF-field and a coil calibration factor for the given Larmor precession frequency. The field-to-current parameter, or calibration parameter, yields the information on how much of the field is changing per unit of current and this has been used to calibrate the response of the RF coil. This method has been repeated across a range of frequencies between 300~Hz to 20~kHz.

\subsection{Field response at different Larmor frequencies }

A quadratic variation in the apparent RF coil calibration with Larmor frequency was observed due to the small, but non-zero, contribution of spin-exchange relaxation to the ground-state relaxation \cite{PhysRevA.16.1877}. \textcolor{black}{The electrical impedance of the RF coil circuit was measured, allowing us to eliminate the coil capacitance or inductance as the underlying reasons for this variation.} Spin-exchange occurs when atoms occupying different hyperfine levels collide in angular-momentum conserving collisions. The sign difference between hyperfine ground states results in rapid dephasing when this process is aggregated across the atomic sample. Our system, utilising repopulation pumping, achieves high inversion between hyperfine ground states, which helps to suppress polarisation relaxation by this process. This is commonly referred to as \textcolor{black}{``light narrowing''} \cite{T_scholtes}. The residual spin exchange broadening causes decoherence within the system, leading to an increase in the transverse relaxation rate $T_{2}$, i.e broadening of the resonance. The spin exchange relaxation rate has been shown to scale as $\Gamma_{SE} \propto \omega_{L}^{2} T$, where $T$ is the spin exchange time \cite{PhysRevA.16.1877}, the rate of spin exchange is quadratic with the Larmor precession frequency. The measured relaxation rate from our resonant response multi fits also shows this squared dependence.
Although spin exchange relaxation is a small contribution to the overall relaxation in our light-narrowed system, there is still a weak correlation between $T_{2}$ and the Larmor frequency. By explicitly including a quadratic contribution to the relaxation rate we can fit this effect in our data analysis and arrive at a robust calibration of the RF coil using the atomic response only.

\subsection {Global fit}

We are able to create a global fit over the whole parameter space of interest. The three independent variables were the RF detuning ($\Delta$), \textcolor{black} {the RF current amplitude ($I_{RF}$)}, and applied static field current ($I_{B0}$). For a given Larmor frequency, the RF frequency is detuned symmetrically around the resonance, followed by an incremental increase in \textcolor{black}{the RF current amplitude}. This detuning process has been repeated for ten distinct \textcolor{black} {RF amplitudes}. After each set of scans, the static field is increased, establishing a new Larmor frequency. This data acquisition procedure is repeated ten times, corresponding to ten different static field values. The dependent variable in each case is the magnitude R of the oscillating atomic response measured on the polarimeter.

\textcolor{black}{The model for the global fit is given below.}  

\begin{equation}
    \textcolor{black}{R = \frac{M_{0}\Omega_{RF}\sqrt{\Delta^{2}+\Gamma^2}}{\Gamma^2+\frac{\Gamma_{1}}{\Gamma_{2}}\Omega_{RF}^2+\Delta^2}}
\end{equation}

\textcolor{black}{Where $\Omega_{RF}$ is the magnetic Rabi frequency ($\Omega_{RF} = \gamma \frac{C_{RF}}{2}I_{RF}$), $\Delta$ is the detuning from the Larmor frequency ($\Delta = \omega_{L} - \omega_{RF}$), $\Gamma_{2}$ is the transverse relaxation rate  ($\Gamma_{2} = \Gamma_{1}+\alpha\omega_{L}$), and $\omega_{L}$ the Larmor frequency ($\omega_{L}$ = $\gamma(C_{B0}I_{B0})+A_{B0}$).}

The fit parameters in the model are: the RF coil calibration parameter $C_{RF}$, static field parameter $C_{B0}$, static field offset $A_{B0}$, spin-exchange factor $\alpha$, longitudinal relaxation rate $\Gamma_{1}$, and the steady state magnetisation $M_{0}$. \textcolor{black}{Figure 4} shows the measured dataset (black) and model fit \textcolor{black} {(pink)}. The complete dataset formed an array with dimensions 50x10x10. Each detuning sweep across the Larmor frequency includes 50 points spanning a resonance width of 180~Hz. \textcolor{black}{The RF current} varies from 20-160~$\mu A$ with equal spacing of 14~$\mu A$, producing ten \textcolor{black}{distinct RF amplitudes} for each Larmor frequency. The Larmor frequency itself is varied from 300~Hz - 20~kHz in  ten steps of 1970~Hz by incrementally increasing the static magnetic field. \textcolor{black}{The x-axis in Fig. 4 denotes the measurement index, this shows how the resonance data changes as a function of the three independent variables. In the case of Fig. 4 (A) this shows a single phase independent magnetic resonance, R, where the RF frequency is detuned around a specific Larmor frequency, at a specific RF field saturation and constant bias field. Fig. 4 (B) shows a collection of R resonances being detuned around a specific Larmor frequency at varying RF saturation. For each resonance the bias field was kept constant, which is equivalent to having the same central Larmor frequency $\omega_{L}$. Fig. 4 (C) shows the entire global fit for varying RF detuning, RF saturation and bias field. Each bias field `column' of data denotes a different Larmor frequency from 300~Hz to 20~kHz ($\omega_{L}(1)$ to $\omega_{L}(10)$) and contained within each `column' is the effect of the RF saturation (Fig. 4(B)) at varying RF detunings (Fig. 4 (A)).}
The main parameter of interest is the RF coil calibration parameter, $C_{RF}$ (nT/mA). \textcolor{black}{This is the calibration factor that pertains to the RF field.} The predicted RF coil calibration based on the size and spacing of the current loops gives \textcolor{black}{36~nT/mA.} The system demonstrated in this work provides a calibration independent of these factors.
For this reason an empirical in-situ calibration is desirable. \textcolor{black}{From the global fitting method, the fitted value for the RF calibration parameter matched that obtained from the geometric calculation (above), as seen in Table 1.} The values of $C_{B0}$ and $A_{B0}$ are found to be consistent with direct measurement of the static field driver, \textcolor{black} {where $C_{B0}$ is the calibration factor of the static field and $A_{B0}$ the static field offset.} From the global fitting method the spin exchange factor is found \textcolor{black}{$\alpha$ = 0.0034 ${Hz}/{Hz^2}$}. \textcolor{black}{$\alpha$ is a phenomenological parameter that parametrises the contribution due to spin exchange as a function of the Larmor frequency} and this implies a small, but non-zero contribution of spin-exchange in the system.  Experimentally, this is expected, as our resonant pumping beam continuously  evacuates atoms from the F=3 ground state. As a result, the majority of the population resides in the stretched F=4 state, and collisions between atoms within the same hyperfine level do not lead to decoherence of their precessing spins. However, the small population that remains in F=3 can exchange spin with the majority population in F=4, and the dephasing due to this contribution becomes increasingly significant with larger Larmor precession frequencies. 
\textcolor{black}{The longitudinal relaxation rate $\Gamma_{1}$ = 5.14 Hz and steady state magnetsiation $M_{0}$ = 0.551 V are found from the fitted model, where the system's pump and probe laser powers, detunings, and the applied RF amplitude vary this response. The relaxation mechanisms of the vapour cell which dictate the minimum magnetic linewidth are well documented independently \cite{Castagna2009}}.

\begin{figure}[h]
    \centering
    \includegraphics[height=0.25\textheight, keepaspectratio]{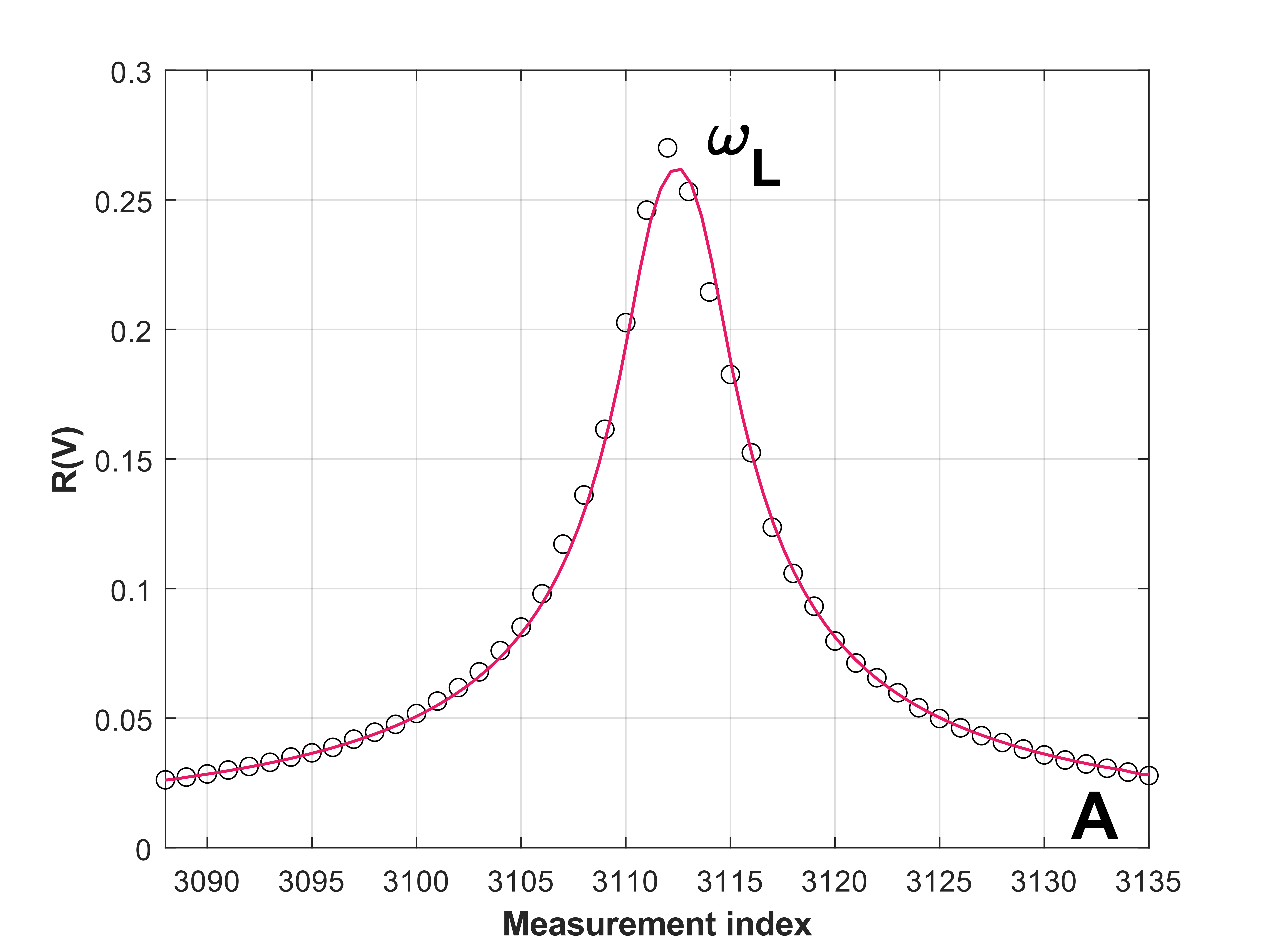} \\
    \vspace{2mm}
    \includegraphics[height=0.25\textheight, keepaspectratio]{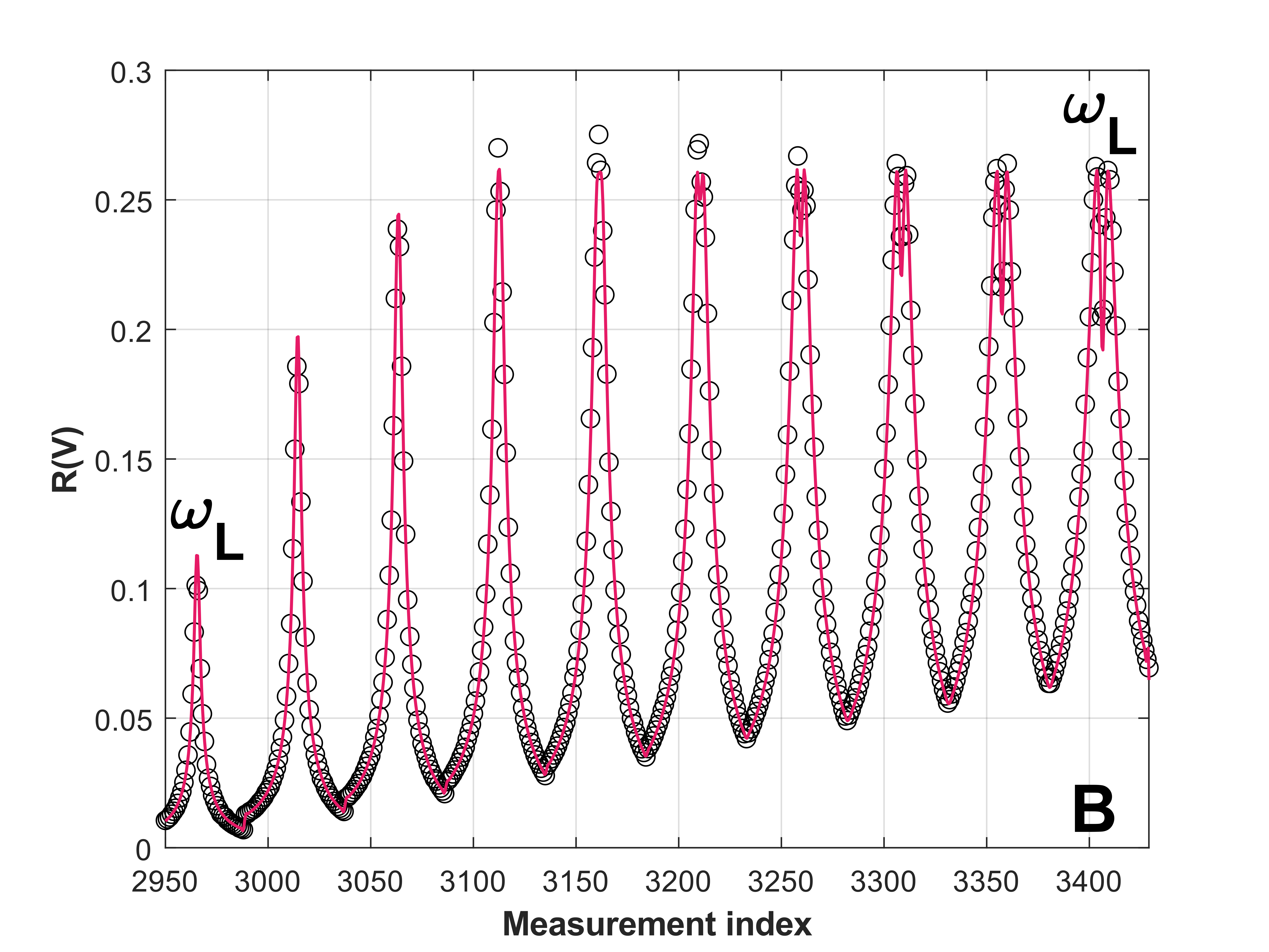} \\
    \vspace{2mm}
    \includegraphics[height=0.25\textheight, keepaspectratio]{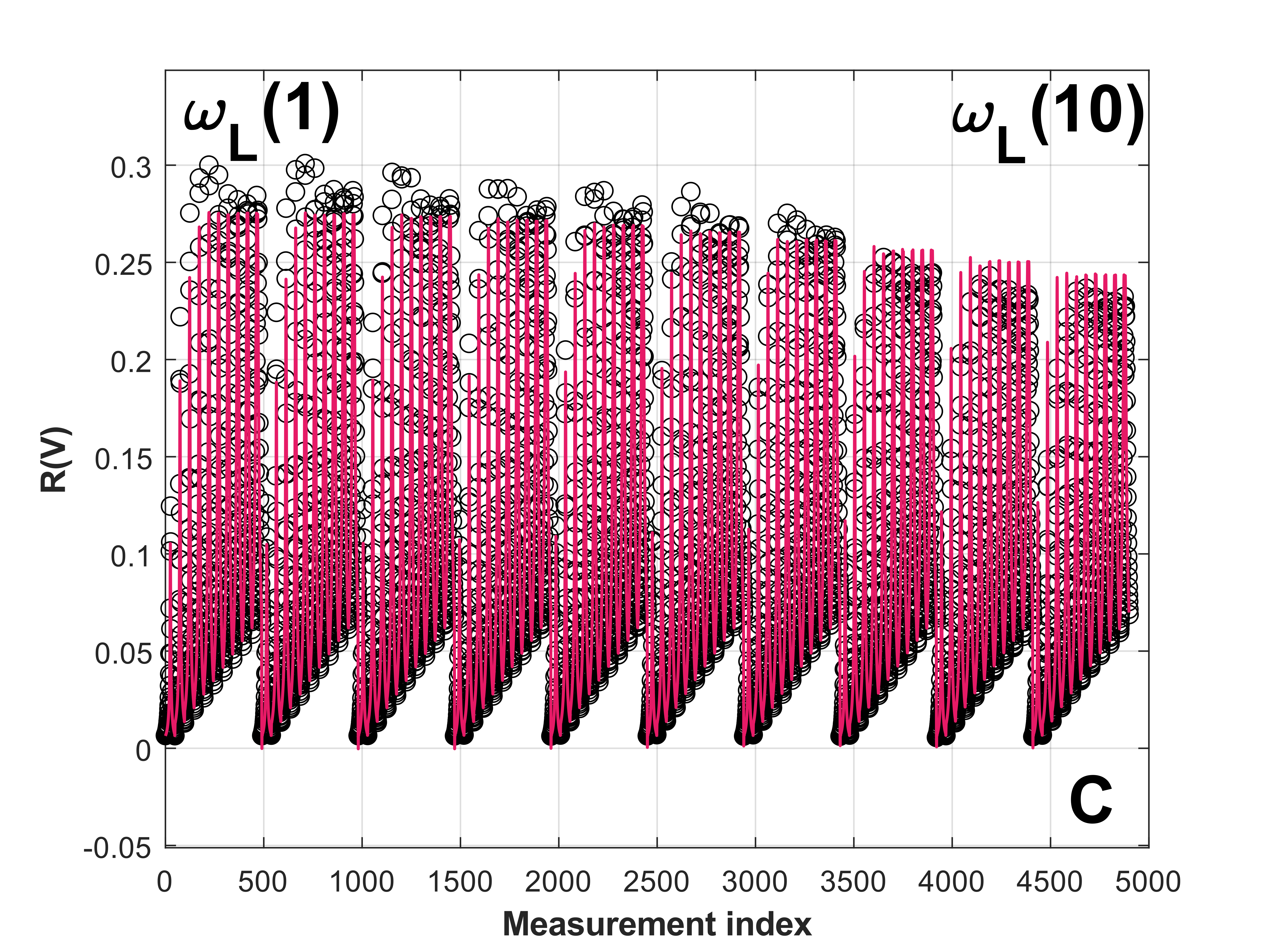}
    \caption{(A) - single resonance of width 180~Hz at a given Larmor frequency. (B) - Ten resonances for a single given Larmor frequency, varying in RF modulation driving depth. (C) - All resonances for varied Larmor frequency and varied RF modulation driving depths.}
\end{figure}

\begin{table}[ht]
\centering
\caption{Table containing fit parameters and absolute errors from the global fitting method}
\begin{tabular}{|c|c|} 
\hline
 $C_{RF}$ & (36.49 $\pm$ 0.09)  nT/mA  \\ 
\hline
 $C_{B0}$ & (65.8041 $\pm$ 0.0003)  nT/mA  \\
\hline
 $A_{B0}$ & (-29.99 $\pm$ 0.02) nT/mA \\
\hline
 $\alpha$ & (0.0034 $\pm$ 0.0001)   Hz/Hz$^{2}$  \\
\hline
 $\Gamma_{1}$ & (5.14 $\pm$ 0.04) Hz \\  
\hline
 $M_{0}$ & ($0.551 \pm 0.001$) V \\ [1ex]
\hline
\end{tabular}
\end{table}

\begin{figure*}[hbt!]
\includegraphics[width=\textwidth]{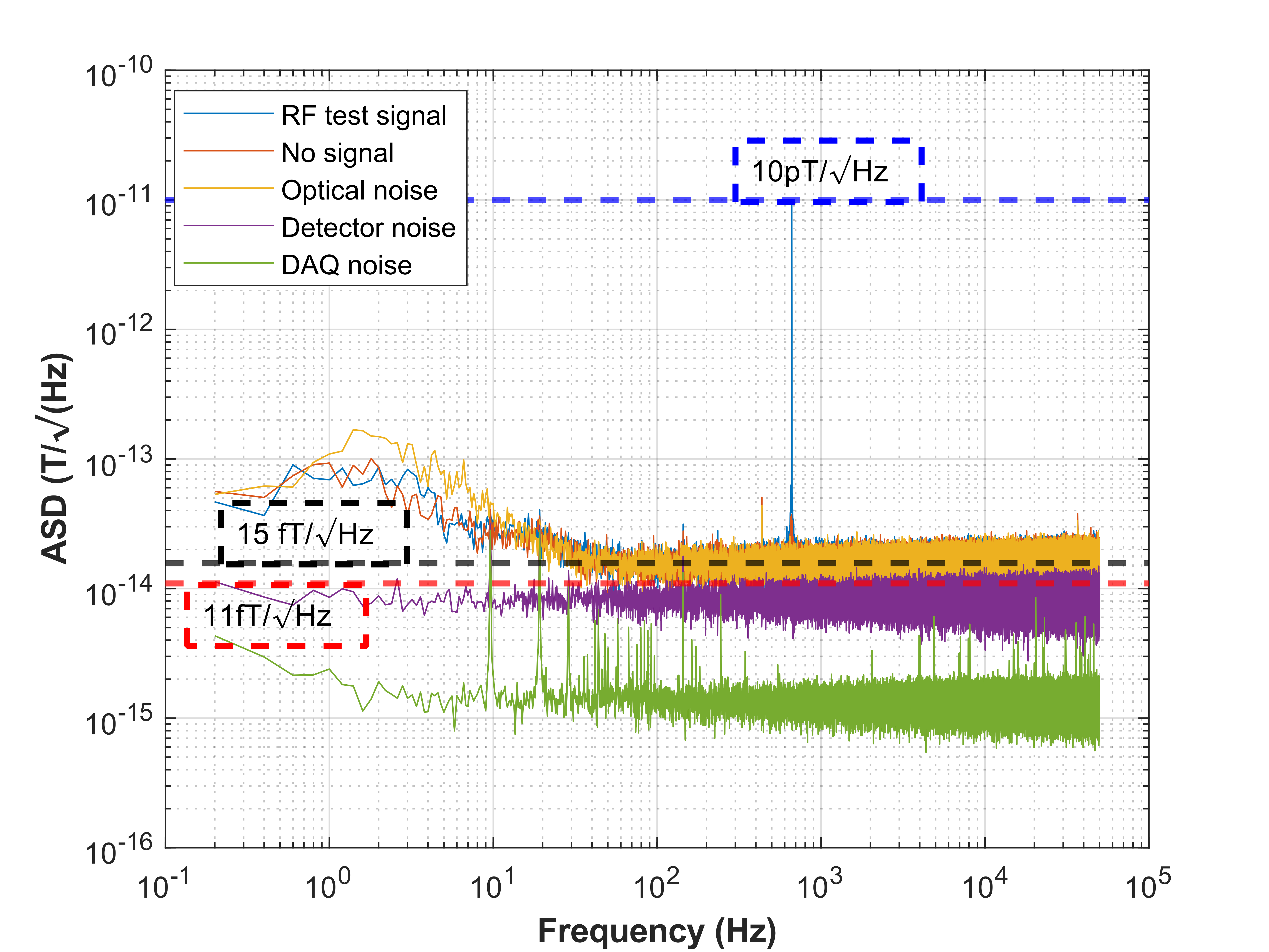}
\centering
\caption{Analysis of different sources of noise in the magnetometer system, each trace denotes a different source of noise and its relative contribution. Operational with test signal of 10~pT (blue), no test signal (orange), optical noise of probe (yellow), detector noise (purple), and electronic DAQ noise (green). The black dashed line represents the average broadband sensitivity in the system, measuring 15~fT/$\sqrt Hz$. The red dashed line is that of the calculated photon shot-noise of 11~fT/$\sqrt Hz$.}
\end{figure*}

\subsection{Calibrated OPM response}
Using the intrinsic calibration procedure discussed previously, the sensitivity of the RF-OPM can be accurately determined by use of a \textcolor{black}{calibrated test modulation}. The performance of the magnetometer is evaluated through a noise analysis of the system. This provided crucial information on the sensors operational noise floors, RF sensitivity, and frequency-dependent factors that limit optimal performance. \textcolor{black}{Figure 5 shows ten averaged amplitude spectral density's (ASDs) for the sensor signal under differing experimental conditions, normalised and calibrated using the global fitting method with a 10~pT field at a Larmor frequency of 667~Hz, shown by the blue trace.}

By removing the RF contribution, the broadband magnetometer noise floor has been measured, corresponding to the ambient magnetic field contributions from the environment (orange trace). Averaging across the bandwidth region dominated by gaussian white noise yielded a broadband noise floor sensitivity of 15~fT/$\sqrt{Hz}$, indicated by the black dashed line in \textcolor{black} {Fig. 5.}
Optical noise, refers to the contribution from the probe beam alone. This is measured by removing the contribution of the pump beam 
(yellow trace). To quantify detector noise, the probe beam is also blocked, leaving the detector with no incident light (purple trace). Finally, the digitiser noise, which represented the intrinsic electronic noise of the DAQ system is measured. A 50~$\Omega$ terminator was connected to the input of the DAQ and the resulting spectrum has been recorded (green trace).

From \textcolor{black}{Fig. 5}, it can be seen that there is little difference in noise contribution across various magnetometer operating states, until the optical noise was removed. This suggests that the optical noise is the limiting factor in the system, and it is close to the photon shot-noise limit. To determine if the system was shot-noise limited at the optical power that yielded optimal magnetometer performance the shot-noise calculation was used, shown below,  

\begin{equation}
\rho_{SN} = G\sqrt{2eP_{det}R}
\end{equation}.

$G$ is the photodetector gain, $e$ the charge of the electron, $P_{det}$ the optical power incident on the detector, and $R$ the responsivity of the detector. \\
In our system, the optimal sensitivity was achieved with a probe power of 125~$\mu W$, at which the calculated photon shot noise is 11~fT/$\sqrt{Hz}$, denoted by the red dashed line in Fig. 6.
By isolating and examining the various sources of noise, it is possible to identify their frequency dependent contributions during different magnetometer operating conditions. DAQ noise shows a range of low-frequency components, resulting from aliased digitiser noise, but these do not exceed the optical noise floor.
Around the Larmor frequency, noise contributions are minimal. The only significant additional signal is from the applied \textcolor{black}{test modulation}. However, below 100~Hz, technical noise became more prominent, rising an order of magnitude above the white noise floor. Unwanted frequencies across the entire bandwidth above 100~Hz contributed minimally to the overall noise. \textcolor{black}{For test modulation frequencies above 100~Hz, the calibration is a robust and reliable, providing an accurate measure of magnetometer sensitivity and overall system performance.}   

\section{conclusion}
A method of absolute field calibration in the ULF and VLF bands are presented, in which measurement of the atomic response to the field was calibrated with respect to the atomic properties of the system. A robust model, in which RF calibration appears as a free parameter, fits well to experimental data, and can be used to provide an RF calibration independent of coil geometry or environment. The method will also operate at carrier frequencies below 300~Hz. Extension below 300 Hz to the SLF band can be achieved by the further work suppressing technical sources of flicker noise.
The inclusion of an empirical parameterisation for spin-exchange relaxation in the atomic system is sufficient to model RF resonance response over the full range of Larmor frequencies. 
Experimentally, this influence was also reduced by increasing the population inversion and employing light narrowing techniques \cite{unknown}. \\
\textcolor{black}{Robust calibration of oscillating fields} is paramount for magnetic metrology applications, particularly in the low-frequency regime. Applications such as low-frequency communications, geophysical surveying, and biomedical imaging require accurate and reliable detection of weak magnetic fields. In magnetic induction tomography (MIT), spatial variations in conductivity are inferred from changes in the magnetic field response of the sample, and the fidelity of these measurements are tied to the systems magnetic field calibration. Similarly, communication systems operating in the ULF and VLF bands - primarily used in attenuating media - must be able to resolve low amplitude magnetic signals with high precision against a noisy background. The calibration technique presented here, which is based solely on the measurement of intrinsic atomic properties and does not rely on an external reference, offers a highly stable and transferable standard for the aforementioned applications.

\begin{acknowledgments}
This work was partly funded by the EPSRC UK Quantum Technology Hubs in Quantum
Enabled Position, Navigation and Timing (QEPNT, EP/Z533178/1), and Quantum
Sensing, Imaging and Timing (QuSIT). The authors would like to thank the group of Prof. emeritus Antoine Weis at the University of Fribourg for the supply of the Cs cell used in this work.
\end{acknowledgments}

\section*{Data Availability Statement}

The data that support the findings of this study are available from the corresponding author upon reasonable request.

\bibliography{aipsamp}

\end{document}